\documentclass[times,sort&compress,3p]{elsarticle}

\usepackage{amssymb}

\usepackage[USenglish]{babel} 
\usepackage{graphicx} 
\usepackage{tikz}
\usepackage{framed}
\usepackage{floatrow}
\usepackage{remreset} 
\usepackage[nouppercase]{scrpage2} 
\usepackage{amsmath} 
\usepackage{amssymb} 
\usepackage[thmmarks, 
amsmath, 
hyperref 
]{ntheorem} 
\usepackage{bm} 
\usepackage{bbm} 
\usepackage{enumitem} 
\usepackage[hang]{subfigure} 
\usepackage{wrapfig} 
\usepackage{tabularx} 
\usepackage{color}
\usepackage{dcolumn} 
\usepackage{booktabs} 
\usepackage{listings} 
\usepackage{psfrag} 
\usepackage[pdftex, 
setpagesize=false, 
pdfborder={0 0 0}, 
pdfpagemode=UseOutlines, 
]{hyperref} 

\theoremstyle{break} 
\theoremheaderfont{\bfseries}
\theorembodyfont{}
\theoremseparator{}
\newtheorem{definition}{Definition} 
\newtheorem{lemma}[definition]{Lemma}

\newtheorem{remark}{Remark} 
\newtheorem{example}{Example} 
\newtheorem{algorithm}{Algorithm}
\theoremstyle{nonumberbreak} 
\theoremsymbol{$\Box$}

\newcommand*{\IP}{\Pr}
\newcommand*{\IE}{{\rm{E}}}

\newcommand*{\IN}{\mathbb{N}}

\journal{Journal of Multivariate Analysis}

\begin{document}

\begin{frontmatter}

\title{Extreme-value copulas associated with the expected scaled maximum \\
of independent random variables}

\author{Jan-Frederik~Mai}
\ead{jan-frederik.mai@xaia.com}
\address{XAIA Investment, Sonnenstr.\ 19, 80331 M\"unchen, Germany}

\begin{abstract}
It is well-known that the expected scaled maximum of non-negative random variables with unit mean defines a stable tail dependence function associated with some extreme-value copula. In the special case when these random variables are independent and identically distributed, min-stable multivariate exponential random vectors with the associated survival extreme-value copulas are shown to arise as finite-dimensional margins of an infinite exchangeable sequence in the sense of De Finetti's Theorem. The associated latent factor is a stochastic process which is strongly infinitely divisible with respect to time, which induces a bijection from the set of distribution functions $F$ of non-negative random variables with finite mean to the set of L\'evy measures $\nu$ on $(0,\infty]$. Since the Gumbel and the Galambos copula are the most popular examples of this construction, the investigation of this bijection contributes to a further understanding of their well-known analytical similarities. Furthermore, a simulation algorithm based on the latent factor representation is developed, if the support of $F$ is bounded. Especially in large dimensions, this algorithm is efficient because it makes use of the De Finetti structure.
\end{abstract}

\begin{keyword}
extreme-value copula \sep De Finetti's Theorem \sep L\'evy measure \sep simulation \sep stable tail dependence function


\end{keyword}

\end{frontmatter}


\section{Introduction}
A $d$-dimensional copula $C$ is a distribution function on $[0,1]^d$ with all one-dimensional margins being uniformly distributed on $[0,1]$. The importance of copulas in multivariate statistics stems from Sklar's Theorem, see \cite{sklar59}, which states that for arbitrary one-dimensional distribution functions $G_1,\ldots,G_d$ the function
$C\{ G_1(t_1),\ldots,G_d(t_d)\}$
(resp.\ $C\{ 1-G_1(t_1),\ldots,1-G_d(t_d)\}$) 
defines a multivariate distribution function (resp.\ survival function) with the pre-defined one-dimensional margins $G_1,\ldots,G_d$. A copula $C$ is of \emph{extreme-value kind} if it satisfies
\begin{gather}
\forall_{t \in (0,\infty)} \; \forall_{u_1, \ldots, u_d \in [0,1]} \quad \{C(u_1,\ldots,u_d)\}^t = C ( u_1^t,\ldots,u_d^t).
\label{EVCproperty}
\end{gather}
This analytical property is usually interpreted in one of the following two ways. 

On one hand, a random vector ${\bf Y}=(Y_1,\ldots,Y_d)$ with survival function defined, for all $t_1,\ldots,t_d \in [0, \infty)$, by
\begin{gather*}
\IP(Y_1>t_1,\ldots,Y_d>t_d)=C( e^{-\lambda_1\,t_1},\ldots,e^{-\lambda_d\,t_d}) 
\end{gather*}
for $\lambda_1,\ldots,\lambda_d \in (0, \infty)$ has a \emph{min-stable multivariate exponential distribution}, which means that the scaled minimum $\min( t_1\,X_1,\ldots,t_d\,X_d )$ is exponentially distributed for all $t_1,\ldots,t_d \in (0, \infty)$; see \cite{esary74}. If one wishes to focus on the dependence structure, it is convenient to normalize the margins to $\lambda_1=\cdots=\lambda_d=1$, which we do henceforth. 

On the other hand, a random vector ${\bf Z}=(Z_1,\ldots,Z_d)$ with distribution function 
\begin{gather}
\IP(Z_1 \leq t_1,\ldots,Z_d \leq t_d)=C \{  G_1(t_1),\ldots,G_d(t_d)\} ,
\label{evd}
\end{gather}
for univariate extreme-value distribution functions $G_1,\ldots,G_d$, has a \emph{multivariate extreme-value distribution}, meaning that it arises as the limit of appropriately normalized componentwise maxima of independent and identically distributed random vectors. If one wishes to focus on the dependence structure, it is convenient to normalize the margins to $G_1(t)=\cdots=G_d(t)=e^{-1/t}$ for all $t \in [0, \infty)$, which we do henceforth. In particular, the distributional relation between ${\bf Y}$ and ${\bf Z}$ after their respective margin normalizations becomes 
$$
{\bf Y}\stackrel{d}{=}1/{\bf Z}, 
$$
with ``$\stackrel{d}{=}$'' denoting equality in distribution. 

For background on extreme-value copulas, the interested reader is referred to \cite{gudendorf09}, and to \cite{nelsen06} for general background on copulas. Due to the defining property (\ref{EVCproperty}) of an extreme-value copula, its so-called \emph{stable tail dependence function,} defined, for all $t_1,\ldots,t_d \in[0, \infty)$, by
\begin{gather}
\ell(t_1,\ldots,t_d) = -\ln\{  C(  e^{-t_1},\ldots,e^{-t_d}) \} 
\label{relation_CF}
\end{gather} 
is homogeneous of order $1$, i.e., $t\times\ell(t_1,\ldots,t_d)=\ell(t\times t_1,\ldots,t\times t_d)$  for all $t \in [0, \infty)$. This property gives rise to a canonical integral representation for the stable tail dependence function, see \cite{dehaan77,ressel13}, given by
\begin{gather}
\ell(t_1,\ldots,t_d) = d\, \IE  \{ \max(t_1\,Q_1,\ldots,t_d\,Q_d) \} ,
\label{Pickandsrepr}
\end{gather}
where the random vector ${\bf Q}=(Q_1,\ldots,Q_d)$ takes values on the unit simplex $S_d \equiv \{{\bf q}=(q_1,\ldots,q_d) \in [0,1]^d : q_1+\cdots+q_d=1\}$, and each component has mean $1/d$. The finite measure $d\IP({\bf Q} \in \mathrm{d}{\bf q})$ on $S_d$ is called the \emph{Pickands dependence measure} associated with $C$, a nomenclature which dates back to \cite{pickands81}.
\par
While the Pickands dependence measure stands in unique correspondence with an extreme-value copula, this does not mean that the stable tail dependence function cannot have an alternative stochastic representation. In particular, if $X_1,\ldots,X_d$ are arbitrary non-negative random variables with unit mean, Segers~\cite{segers12} showed that setting, for all $t_1,\ldots,t_d \in [0, \infty)$, 
\begin{gather*}
\ell(t_1,\ldots,t_d) \equiv  \IE \{  \max (t_1\,X_1,\ldots,t_d\,X_d) \} ,
\end{gather*}
defines a proper stable tail dependence function of some extreme-value copula, which yields a useful construction device for parametric models. In the present article, we study the associated extreme-value copulas in the special case when $X_1,\ldots,X_d$ are independent. Denoting their distribution functions by ${\bf F}=(F_1,\ldots,F_d)$, we denote, for all $t_1,\ldots,t_d \in [0, \infty)$, 
\begin{gather}
\ell_{\bf F}(t_1,\ldots,t_d) \equiv  \IE\{ \max (t_1\,X_1,\ldots,t_d\,X_d) \} ,
\label{stabletaib_F}
\end{gather}
and the extreme-value copula associated with $\ell_{\bf F}$ via (\ref{relation_CF}) is denoted by $C_{\bf F}$. 

The main contribution of the present article is a detailed study of the De Finetti structure of $C_{\bf F}$ in the special case when $F_1=\cdots=F_d = F$. The computations in \cite{dombry16} point out that the two most prominent representatives in this family of extreme-value copulas are the Gumbel copula ($F$ is a certain Fr\'echet distribution) and the Galambos copula ($F$ is a certain Weibull distribution). The Gumbel copula is named after Emil Gumbel \cite{gumbel60,gumbel61}, whereas the Galambos copula is named after J\'anos Galambos~\cite{galambos75}. Moreover, the recent articles \cite{genest17,belzile17} point out some further striking similarities between the Gumbel and the Galambos extreme-value copulas.  
\par
The remainder of the article is organized as follows. Section~\ref{sec_definetti} considers the case when $F_1=\cdots=F_d=F$, in which case we also write $\ell_{\bf F}=\ell_{F}$ and $C_{\bf F}=C_F$. An infinite exchangeable sequence $(Y_k)_{k \in \IN}$ of random variables is constructed such that for each integer $d \in \mathbb{N}$, the random vector $(Y_1,\ldots,Y_d)$ has a min-stable multivariate exponential distribution with associated stable tail dependence function $\ell_F$. It follows that the conditional cumulative hazard process $H_t\equiv -\ln \{ \IP(Y_1>t\,|\,\mathcal{H})\} $ is strongly infinitely divisible with respect to time in the sense of \cite{maischerer13}, where $\mathcal{H}$ denotes the tail-$\sigma$-field of $(Y_k)_{k \in \IN}$ in the sense of De Finetti's Theorem; see \cite{definetti31,definetti37,aldous85}. 
The relation between the associated L\'evy measure $\nu_F$ on $(0,\infty]$ and the distribution function $F$ is explored. 

Section~\ref{sec_simu} enhances the stochastic model to allow for the non-exchangeable case of arbitrary $F_1,\ldots,F_d$. In particular, the De Finetti construction of the preceding section is slightly enhanced to derive a similar stochastic model for a min-stable multivariate exponential random vector $(Y_1,\ldots,Y_d)$ with stable tail dependence function $\ell_{\bf F}$. It is based on $d$ latent frailty processes $(H_t^{(1)})_{t \geq 0}, \ldots, (H_t^{(d)})_{t \geq 0}$ which are dependent. Simulation algorithms for the new family are discussed. If the supports of $F_1,\ldots,F_d$ are all bounded, the aforementioned frailty model can be used for exact simulation. The latent frailty processes on which this simulation algorithm is based, resemble shot-noise processes in this case. In the general case of possibly unbounded supports of $F_1,\ldots,F_d$, an exact simulation strategy of \cite{dombry16}, based on the Pickands dependence measure, can be applied. In particular, the simulation of ${\bf Q}$ in Eq.~\eqref{Pickandsrepr} is straightforward for the family of extreme-value copulas $C_{\bf F}$. Section~\ref{sec_conc} concludes.    

\section{Stochastic construction as infinite exchangeable sequence}\label{sec_definetti}

Let $F$ be the distribution function of a non-negative random variable with finite mean and $F(0)<1$ (the random variable is not identically zero), and $(\epsilon_k)_{k \in \IN}$ a sequence of independent and identically distributed random variables with unit exponential distribution. Throughout, in order to include boundary cases and simplify notation, we define $1/0\equiv \infty$, $F(\infty)\equiv 1$, $-\ln (0)\equiv \infty$, and $e^{-\infty} \equiv 0$. Concerning further notation, throughout the article we denote by $\delta_x$ the Dirac measure at $x \in (0,\infty)$. Furthermore, for each $t \in [0,1]$, we denote
\begin{gather*}
F^{-1}(t)\equiv \inf\{x>0 : F(x) \geq t\}
\end{gather*}
the \emph{generalized inverse} of the distribution function $F$ at $t$, and set $b_F\equiv F^{-1}(0)$, and $u_F\equiv F^{-1}(1)$, the lower and upper end points of the support of $F$, respectively. 
\par
Denoting by $F(x-)\equiv \lim_{t \uparrow x}F(t)$ the left-continuous version of the (right-continuous) distribution function $F$, we consider the stochastic process defined, for all $t \in [0, \infty)$, by
\begin{gather*}
H_t\equiv -\ln \left\{\prod_{k=1}^{\infty}\,F \left(  \frac{\epsilon_1+\cdots+\epsilon_k}{t}- ~\right)  \right\},
\end{gather*}  
which takes values in $[0,\infty]$. By definition, $H_0=0$, and $t \mapsto H_t$ is almost surely right-continuous and non-decreasing. Also, $\lim_{t \rightarrow \infty}H_t=\infty$, which is obvious if $F(0)=0$. For $F(0)>0$ we have by assumption that $F(0)<1$, and thus also $\lim_{t \rightarrow \infty}H_t=\infty$. The following lemma shows in particular that the infinite product in the definition of $H_t$ converges with positive probability. In fact, if $b_F=0$ it even converges with probability~$1$.

\begin{lemma}[Laplace transform of $H_t$] \label{lemma_welldef}
The Laplace transform of the random variable $H_t$ is given, for all $u \in [0, \infty)$, by
\begin{gather*}
\IE (  e^{-u\,H_t})  = e^{-t\,\Psi_F(u)}\quad \mbox{with} \quad \Psi_F(u)\equiv \int_{0}^{\infty}\{ 1-F(x)^u\} \,\mathrm{d}x,
\end{gather*}
and satisfies $\IP(H_t<\infty) = e^{-t\,b_F}$ for all $t \in [0, \infty)$.
\end{lemma}

\bigskip
\noindent
\textit{Proof.}
Define $P\equiv \sum_{k=1}^{\infty}\delta_{\epsilon_1+\cdots+\epsilon_k}$, so that $P$ is a Poisson random measure with mean measure $\mathrm{d}x$. Resorting to the Laplace functional formula of Poisson random measure, see \cite[Proposition 3.6]{resnick87}, the Laplace transform of $H_t$ is given by
\begin{align*}
\IE (  e^{-u\,H_t})  &=\IE \Big[  e^{-\int_{0}^{\infty}-\ln \{  F(x/t-)^u\} \,P(\mathrm{d}x)} \Big]  = \exp\left[  -\int_{0}^{\infty}\{ 1-F(x/t)^u\} \,\mathrm{d}x \right]  = \exp \left[  -t\,\int_{0}^{\infty}\{ 1-F(x)^u\} \,\mathrm{d}x \right] = e^{-t\,\Psi_F(u)}.
\end{align*}
For $u\in [0,1]$ we have that $1-F(x)^u \leq 1-F(x)$, which is integrable from the assumption that $F$ is the distribution function of a random variable with finite mean $\int_0^{\infty} \{ 1-F(x)\} \,\mathrm{d}x<\infty$. This allows one to apply  Lebesgue's dominated convergence theorem in $(\ast)$ below to obtain
\begin{align*}
\IP(H_t<\infty) &= \IE \{ \mathbf{1}_{(H_t<\infty)}\} =\IE \left(  \lim_{u \downarrow 0}\, e^{-u\,H_t} \right) =\lim_{u \downarrow 0}\IE \big(  e^{-u\,H_t} \big)   \\
& =  \exp\left[  -t\,\lim_{u \downarrow 0}\int_{0}^{\infty}\{ 1-F(x)^u\} \,\mathrm{d}x \right]   \stackrel{(\ast)}{=} \exp\left[  -t\,\int_{0}^{\infty} \{  1- \mathbf{1}_{(x>b_F)}\} \,\mathrm{d}x \right]  =e^{-t\,b_F},
\end{align*}
establishing the claim. Notice that the bounded convergence theorem has been applied in the third equality. \hfill $\Box$

\bigskip
Recall that a function $\Psi:[0,\infty) \rightarrow [0,\infty)$ is called \emph{Bernstein function} if $\Psi(0)=0$, $\Psi$ is infinitely often differentiable on $(0,\infty)$, with a possible jump at zero, and its first derivative is completely monotone on $(0,\infty)$; see \cite{schilling10} for background on these. A Bernstein function has a canonical representation of the form
\begin{gather}
 \Psi(u) = \mu\,u+\int_{(0,\infty)}( 1-e^{-u\,x} ) \,\nu(\mathrm{d}x)+\nu(\{\infty\})\,\mathbf{1}_{\{u>0\}}=\mu\,u+\int_{(0,\infty]}( 1-e^{-u\,x} ) \,\nu(\mathrm{d}x)
\label{BF_repr}
\end{gather}
with $u \in [0, \infty)$ and $\nu$ a Radon measure  on $(0,\infty]$ satisfying the integrability condition
\begin{gather}
\int_0^{1}x\,\nu(\mathrm{d}x)<\infty, 
\label{levycondition}
\end{gather}
called the \emph{L\'evy measure} of $\Psi$, and a \emph{drift constant} $\mu \geq 0$. The number $\nu(\{\infty\})$ is called the \emph{killing rate} of $\Psi$, and the so-called L\'evy--Khintchine representation (\ref{BF_repr}) gives a one-to-one relationship between Bernstein functions and pairs $(\mu,\nu)$ of drift constants and L\'evy measures. By well-known results from the theory on infinite divisibility, it already follows from Lemma~\ref{lemma_welldef} that $\Psi_F$ is a Bernstein function --- whose associated L\'evy measure will be examined below in Lemma~\ref{lemma_levy} --- and that $(H_t)_{t \geq 0}$ is weakly infinitely divisible with respect to time, meaning that there exists a (possibly killed) L\'evy subordinator $(L_t)_{t \geq 0}$ such that $L_t \stackrel{d}{=}H_t$ for all $t \in [0, \infty)$. For background on L\'evy subordinators the interested reader is referred to the textbooks \cite{bertoin96,sato99}. 

By virtue of Theorem~5.3 in \cite{maischerer13}, the next lemma shows that $(H_t)_{t \geq 0}$ is even strongly infinitely divisible with respect to time, meaning that
\begin{gather*}
\forall_{n \in \mathbb{N}} \quad (H_t)_{t \geq 0} \stackrel{d}{=} (H^{(1)}_{t/n}+\cdots+H^{(n)}_{t/n})_{t \geq 0},
\end{gather*}
where $(H^{(1)}_t),\,(H^{(2)}_t),\ldots$ are independent copies of $H_t$. We denote by $(\xi_k)_{k \in \IN}$ an independent copy of $(\epsilon_k)_{k \in \IN}$ and define the infinite exchangeable sequence of random variables $(Y_k)_{k \in \IN}$, where, for each $k \in \IN$,
\begin{gather}
Y_k\equiv \inf\{t>0 : H_t > \xi_k\}.
\label{definetti_model}
\end{gather}

\begin{lemma}[De Finetti construction] \label{lemma_DeFinetti}
Assume that $F$ has unit mean. For arbitrary $d \in \mathbb{N}$ the random vector $(Y_1,\ldots,Y_d)$, as defined in (\ref{definetti_model}), has a min-stable multivariate exponential distribution with survival function defined, for all $t_1,\ldots,t_d \in [0, \infty)$, by
$
\IP(Y_1>t_1,\ldots,Y_d>t_d) = \exp\{ -\ell_{F}(t_1,\ldots,t_d)\} ,
$
where the function $\ell_{F}$ is given by (\ref{stabletaib_F}) with $X_1,\ldots,X_d$ independent and identically distributed with distribution function~$F$.
\end{lemma}

\bigskip
\noindent
\textit{Proof.}
As in Lemma \ref{lemma_welldef}, let $P\equiv \sum_{k=1}^{\infty}\delta_{\epsilon_1+\cdots+\epsilon_k}$ be a Poisson random measure with mean measure $\mathrm{d}x$. For $t_1,\ldots,t_d \in [0, \infty)$ we compute similar as in Lemma \ref{lemma_welldef} that
\begin{align*}
\IP(Y_1>t_1,\ldots,Y_d>t_d)  = \IE \big \{ e^{-(H_{t_1}+\cdots+H_{t_d})} \big \}  = \IE \Big[  e^{-\int_{0}^{\infty}-\ln \{  \prod_{i=1}^{d}F(x/t_i-)\} \,P(\mathrm{d}x)} \Big]  = \exp\left[  -\int_{0}^{\infty}\Big\{ 1-\prod_{i=1}^{d}F(x/t_i) \Big\} \,\mathrm{d}x \right] .
\end{align*}
Furthermore,
\begin{align*}
\int_{0}^{\infty}\Big\{ 1-\prod_{i=1}^{d}F(x/t_i) \Big\} \,\mathrm{d}x = \int_{0}^{\infty}\IP\{ \max ( t_1\,X_1,\ldots,t_d\,X_d) >x\}  \,\mathrm{d}x =  \IE \{  \max(t_1\,X_1,\ldots,t_d\,X_d)\}.
\end{align*}
This completes the argument. \hfill $\Box$

\bigskip
Conditioned on the $\sigma$-algebra $\mathcal{H}$ generated by the path of $(H_t)_{t \geq 0}$, which coincides almost surely with the tail-$\sigma$-field of the sequence $(Y_k)_{k \in \IN}$, see Corollary~3.12 in \cite{aldous85}, the random variables $(Y_k)_{k \in \IN}$ are independent and identically distributed with distribution function given, for all $t \in [0, \infty)$, by
\begin{gather*}
1-e^{-H_t} = 1-\prod_{k\geq 1}F\left(  \frac{\epsilon_1+\cdots+\epsilon_k}{t}- \right) ,
\end{gather*}
and with conditional cumulative hazard process $(H_t)_{t \geq 0}$. Such conditional hazard processes associated with min-stable multivariate exponential distributions have a close relationship with the concept of infinite divisibility, as explored in \cite{maischerer13}. In particular, as already mentioned, the function $\Psi_F$ is a Bernstein function and as such it has a L\'evy--Khintchine representation given, for all $u \in [0, \infty)$, by 
\begin{gather}
\Psi_F(u) = \int_{(0,\infty]}(  1-e^{-u\,x}) \,\nu_F(\mathrm{d}x),
\label{levybernstein}
\end{gather}
with zero drift $\mu=0$ and some L\'evy measure $\nu_F$. For an arbitrary L\'evy measure $\nu$ on $(0,\infty]$ we denote by $S_{\nu}(t)\equiv \nu(  (x,\infty]) $ and
\begin{gather*}
S_{\nu}^{-1}(t) \equiv  \inf\big\{ x>0 : S_{\nu}(x) \leq t\big\},\quad t \in [ \nu(\{\infty\}),\nu((0,\infty])] ,
\end{gather*}
its associated survival function and the related generalized inverse thereof. Recall in particular that the function $S_{\nu}$ determines the measure $\nu$.

\begin{lemma}[The associated L\'evy measure] \label{lemma_levy}
The L\'evy measure $\nu_F$ associated with the distribution function $F$ via (\ref{levybernstein}) is determined by its survival function given, for all $t \in [0, \infty)$, by
\begin{gather}
S_{\nu_F}(t)= F^{-1}(  e^{-t}) .
\label{levycorrespondence}
\end{gather}
Furthermore, the mapping  $F \mapsto \nu_F$ from distributions on $[0,\infty)$ with finite, positive mean to the set of L\'evy measures on $(0,\infty]$ is a bijection. The inverse function $\nu \mapsto F_{\nu}$ assigns to a L\'evy measure $\nu$ the distribution function
\begin{gather*}
F_{\nu}(t)\equiv \begin{cases}
0 & \mbox{if }t<\nu(\{\infty\}),\\
e^{-S_{\nu}^{-1}(t)} & \mbox{if }\nu(\{\infty\}) \leq t<\nu( (0,\infty] ), \\
1 & \mbox{if }t \geq \nu( (0,\infty] ) .
\end{cases}
\end{gather*}
\end{lemma}

\bigskip
\noindent
\textit{Proof.}
First of all, we observe as a consequence of the right-continuity of $F$ that
\begin{gather}
\forall_{x \in (0,1]} \; \forall_{t \in (0,u_F)} \quad F(t)<x \;\Leftrightarrow \; t<F^{-1}(x).
\label{geninv}
\end{gather}
Denoting by $\lambda$ the Lebesgue measure on $(0,\infty)$, we observe that the map $G\equiv -\ln (F):(0,u_F)\rightarrow (0,\infty]$ is measurable. Consider the measure $G_{\lambda}$ defined by $G_{\lambda}(E)\equiv \lambda \{G^{-1}(E)\}$, $E$ a Borel set in $(0,\infty]$, where $G^{-1}(E)$ denotes the pre-image of the set $E$. Then we observe for $x \in (0,\infty]$ that
\begin{align*}
G_{\lambda}(  (x,\infty])  &= \lambda[  \{t \in (0,u_F) :  -\ln \{F(t)\} > x\}] = \lambda[  \{t \in (0,u_F) :  F(t) < e^{-x}\}] \\
& \stackrel{(\ref{geninv})}{=}\lambda[  \{t \in (0,u_F) : t < F^{-1}(e^{-x}) \}]  = F^{-1}(e^{-x}) = \nu_F((x,\infty]).
\end{align*}
Consequently, $G_{\lambda}=\nu_F$, establishing the measure-theoretic change of variable formula
\begin{gather}
\int_{(0,\infty]}g(x)\,\nu_F(\mathrm{d}x) = \int_{(0,u_F)}g\{ G(x) \} \,\lambda(\mathrm{d}x)=\int_{(0,u_F)}g[ -\ln \{F(x)\}]  \,\mathrm{d}x,
\label{changeofvar}
\end{gather}
for measurable functions $g(x)$. Plugging in the function $g(x)=1-\exp(-u\,x)$, this implies
\begin{align*}
\Psi_F(u) = \int_{0}^{\infty}\{ 1-F(x)^u\} \,\mathrm{d}x=\int_{(0,u_F)} \left[ 1-e^{-u\,[-\ln \{F(x)\}]}\right] \,\mathrm{d}x \stackrel{(\ref{changeofvar})}{=} \int_{(0,\infty]}( 1-e^{-u\,x}) \,\nu_F(\mathrm{d}x).
\end{align*}
Furthermore, since $F$ has finite mean, it follows from the last equality that
\begin{align*}
\int_{0}^{1}x\,\nu_F(\mathrm{d}x) \leq 2\,\int_{0}^{1}( x- {x^2}/{2}) \,\nu_F(\mathrm{d}x)\leq 2\,\int_{(0,\infty]}( 1-e^{-x}) \,\nu_F(\mathrm{d}x) =2\, \int_{0}^{\infty}\{ 1-F(x)\} \,\mathrm{d}x<\infty,
\end{align*}
so $\nu_F$ satisfies the integrability condition (\ref{levycondition}), hence is a proper L\'evy measure. In order to verify that $F \mapsto \nu_F$ is a bijection, it suffices to check\ldots
\begin{itemize}
\item[(a)]  for a distribution function $F$ with finite mean that $F_{\nu_F}=F$;
\item[(b)] for a L\'evy measure $\nu$ that $\nu_{F_{\nu}}=\nu$. 
\end{itemize}
To see (a), let $t \in [b_F,u_F)$ arbitrary, and observe that $b_F=\nu_F(\{\infty\})$ and $u_F=\nu_F((0,\infty])$ by definition. Further,
\begin{gather}
\inf\{x>0 : F^{-1}(e^{-x}) \leq t\} = -\ln \{F(t)\}.
\label{infshortref}
\end{gather}
To verify (\ref{infshortref}), denote $I\equiv \inf\{x>0 : F^{-1}(e^{-x}) \leq t\}$. Obviously, $I \leq -\ln \{F(t)\}$, so that $\exp(-I) \geq F(t)$. Now we assume there exists $\delta \in (0, \infty)$ such that $\exp(-I) \geq F(t)+\delta$ and derive a contradiction. This assumption implies that $I \leq -\ln \{F(t)+\delta\}$. By definition of the infimum this implies that $F^{-1}\big\{ F(t)+\delta \big\} \leq t$, which is clearly a contradiction, so (\ref{infshortref}) is valid. Consequently,
\begin{align*}
 F_{\nu_F}(t) = e^{-S_{\nu_F}^{-1}(t)} = \exp \big[  -\inf\{x>0 : F^{-1}(e^{-x}) \leq t\}\big]  \stackrel{(\ref{infshortref})}{=} F(t),
\end{align*}
establishing (a). To see (b), we need to show for $x \in (0, \infty)$ that $F_{\nu}^{-1}( e^{-x}) =S_{\nu}(x)$. To this end,
\begin{align*}
F_{\nu}^{-1}( e^{-x}) =\inf\{t>0 : F_{\nu}(t) \geq e^{-x}\}=\inf\{t>0 : e^{-S_{\nu}^{-1}(t)} \geq e^{-x}\} = \inf\{t>0 : S_{\nu}^{-1}(t) \leq x\} = S_{\nu}(x),
\end{align*}
where the last equality holds, since for arbitrary $\epsilon \in [0, \infty)$ it is observed that $S_{\nu}^{-1}\{S_{\nu}(x)+\epsilon \}\leq S_{\nu}^{-1} \{S_{\nu}(x)\} \leq x$. Finally, we check that the definition of $F_{\nu}$ gives a distribution function with finite mean. We have already seen that there is a unique distribution function $F$ with finite mean such that $\nu=\nu_F$. But we have also seen that $\nu=\nu_{F_{\nu}}$, so that $F_{\nu}=F$ has finite mean. \hfill $\Box$

\bigskip
\begin{remark}[A subtle technicality]
Both non-increasing functions $S_{\nu}$ and $S_{\nu}^{-1}$ are right-continuous, explaining the right-continuity of $F_{\nu}$, which is defined as a continuous function of the right-continuous function $S_{\nu}^{-1}$. Right-continuity of $S_{\nu}$ is clear by definition, and right-continuity of $S_{\nu}^{-1}$ can be shown completely analogous to Proposition 2.3(2) in \cite{embrechts13}. This is a subtle difference to the case of generalized inverses of non-decreasing functions. For instance, $F^{-1}$ is left-continuous for the right-continuous distribution function $F$, see \cite[Proposition 2.3(2)]{embrechts13}. Further, since $t \mapsto \exp(-t)$ is decreasing, $t \mapsto F^{-1}\{\exp(-t)\}$ is right-continuous, which explains the correctness of (\ref{levycorrespondence}).
\end{remark}

It follows from Lemma 2.15 and Corollary 3.12 in \cite{aldous85} that the probability distribution of $(Y_k)_{k \in \IN}$ is uniquely determined by that of $(H_t)_{t \geq 0}$, and vice versa. Since two infinitely divisible distributions with different L\'evy measures are truly different, Lemma \ref{lemma_levy} implies that two different distribution functions $F_1$ and $F_2$ induce two truly different extreme-value copulas $C_{F_1}$ and $C_{F_2}$. Furthermore, the stable tail dependence function may alternatively be written in terms of the L\'evy measure $\nu_F$, which amounts to
\begin{gather*}
\ell_F(t_1,\ldots,t_d) = \int_{(0,\infty]} \Big\{ 1-e^{-\sum_{i=1}^{d}S_{\nu_F}^{-1}( x/t_i)} \Big \} \,\mathrm{d}x.
\end{gather*}
It is educational to remark that existence of the mean of $F$ corresponds to the integrability condition (\ref{levycondition}) on the level of the associated L\'evy measure $\nu_F$, and that $\Psi_F(1)$ equals the mean of $F$. Furthermore, the L\'evy measure $\nu_F$ is finite if and only if the support of $F$ is bounded, and $\nu_F(\{\infty\})=b_F$ as well as $\nu_F((0,\infty])=u_F$. An atom of $F$ at zero, i.e., $F(0)>0$, corresponds to bounded support of the L\'evy measure, since we see
\begin{gather*}
F(0) = \exp \big[ -\inf\big\{x>0 : \nu_F(  (x,\infty]) =0\big\} \big] .
\end{gather*}
Finally, absolute continuity of $F$ translates to absolute continuity of the L\'evy measure $\nu_F$, as the following remark points out.

\bigskip
\begin{remark}[Special case of absolutely continuous distributions]
Under the bijection of Lemma \ref{lemma_levy}, distribution functions $F$ with positive density $f_F$ on $(0,\infty)$ correspond to L\'evy measures $\nu$ with positive density $f_{\nu}$ on $(0,\infty)$, and the bijection boils down to the density transformation formulas
\begin{align*}
f_{\nu_F}(x) = \frac{e^{-x}}{f_F \{  F^{-1}(e^{-x})\} },\quad f_{F_{\nu}}(x)=\frac{e^{-S_{\nu}^{-1}(x)}}{f_{\nu}\{  S_{\nu}^{-1}(x)\} },
\end{align*}
where $S_{\nu}^{-1}$ is the (regular) inverse of the function $x \mapsto \nu(  (x,\infty)) $. 
\end{remark}

Examples \ref{ex_stable} and \ref{ex_galambos} below demonstrate how the considered family of extreme-value copulas comprises both the Gumbel and the Galambos copula as well-known special cases.

\begin{example}[The Gumbel copula]\label{ex_stable}
This family is parameterized by $\theta \in (0,1)$. The L\'evy measure is $\nu(\mathrm{d}x)\equiv \theta\,x^{-1-\theta}/\Gamma(1-\theta)\,\mathrm{d}x$, with associated distribution function defined, for all $t \in [0, \infty)$, by
\begin{gather}
F(t) = \exp[  - \{ \Gamma(1-\theta)\,t\} ^{- {1}/{\theta}}] ,
\label{gumbeldf}
\end{gather}
which is the Fr\'echet distribution with shape parameter $1/\theta$ and unit mean. Hence, for all $t \in [0, \infty)$,
\begin{gather*}
H_t = t^{ {1}/{\theta}}\,\Gamma(1-\theta)^{- {1}/{\theta}}\,\sum_{k \geq 1}(\epsilon_1+\cdots+\epsilon_k)^{- {1}/{\theta}}.
\end{gather*}
In particular, Lemma~3.3 in \cite{maischerer13} implies that the random variable $\Gamma(1-\theta)^{- {1}/{\theta}}\,\sum_{k \geq 1}(\epsilon_1+\cdots+\epsilon_k)^{-{1}/{\theta}}$ is $\theta$-stable, i.e., has Laplace transform $\varphi_{\theta}(u)\equiv \exp(-u^{\theta})$, see also Section 4.2 in \cite{bondesson82}. The survival function of $(Y_1,\ldots,Y_d)$ is given by
\begin{gather*}
\IP(Y_1>t_1,\ldots,Y_d>t_d) = \varphi_{\theta}\{ \varphi_{\theta}^{-1}(e^{-{t_1}})+\cdots+\varphi_{\theta}^{-1}(e^{-{t_d}})\} ,
\end{gather*}
i.e., has survival copula of Archimedean kind with generator equal to the Laplace transform $\varphi_{\theta}$; see \cite{neilnev09} for background on Archimedean copulas. This is the so-called Gumbel copula. Genest and Rivest~\cite{genest89} were the first to observe that the Gumbel copula is the only copula which is both Archimedean and of extreme-value kind. Based on nice algebraic properties of the involved stable distribution, several asymmetric generalizations of the Gumbel copula model have been derived and applied to real-world data in the literature. Prominent examples include \cite{fougeres09,reich12}.  
\end{example}

\begin{remark}[Alternative representation of $\ell_F$]\label{rmk_alt}
Applying the principle of inclusion and exclusion, it is readily verified that 
\begin{align*}
\ell_{\bf F}(t_1,\ldots,t_d) &= \IE \{ \max (t_1\,X_1,\ldots,t_d\,X_d )\}  =\sum_{k=1}^{d}(-1)^{k+1}\sum_{1 \leq i_1<\cdots<i_k \leq d}\IE \{ \min (t_{i_1}\,X_{i_1},\ldots,t_{i_k}\,X_{i_k}) \} \\
& =\sum_{k=1}^{d}(-1)^{k+1}\sum_{1 \leq i_1<\cdots<i_k \leq d}\int_0^{\infty}\prod_{j=1}^{k}\big\{1-F_{i_j}(  {x}/{t_{i_j}}) \big\}\,\mathrm{d}x.
\end{align*}
This alternative representation might be advantageous if $F_1,\ldots,F_d$ are such that expected scaled minima are easier to compute than expected scaled maxima, i.e., if the survival functions rather than the distribution functions have an analytical form that is better compatible with products.
\end{remark}

A prominent application of the representation in Remark \ref{rmk_alt} is the Galambos copula.

\bigskip
\begin{example}[The Galambos copula]\label{ex_galambos}
This parametric family is parameterized by $\theta \in (0,\infty)$. The L\'evy measure is $\nu(\mathrm{d}x)\equiv e^{-x}/(1-e^{-x})\,\{ -\ln (1-e^{-x})\} ^{\theta-1}/\Gamma(\theta)\,\mathrm{d}x$, and has been investigated in \cite{mai14}. The associated distribution  function given, for all $t \in [0, \infty)$, by
\begin{gather}
F(t)= 1-e^{-\{  t\,\Gamma(\theta+1)\} ^{ {1}/{\theta}}},
\label{galambosdf}
\end{gather}
is the Weibull distribution with shape parameter $1/\theta$ and scale parameter $1/\Gamma(\theta+1)$. Making use of Remark~\ref{rmk_alt}, the resulting stable tail dependence function is
\begin{align*}
\ell_{F}(t_1,\ldots,t_d) = \sum_{j=1}^{d}\,(-1)^{j+1} \,\sum_{1 \leq i_1<\cdots<i_j \leq d}\left(  \sum_{k=1}^{j}t_{i_k}^{-\theta} \right) ^{- {1}/{\theta}},
\end{align*}
and $C_F$ is the so-called Galambos copula, named after \cite{galambos75}. Genest et al.\ \cite{genest18} embed the Galambos copula into a larger family of copulas termed reciprocal Archimedean copulas, and point out that the Galambos copula is the only copula which is both reciprocal Archimedean and of extreme-value kind.
\end{example}

\bigskip
\begin{example}[Upper Fr\'echet bound]
In the very special case $F(t)\equiv \mathbf{1}_{(t \geq 1)}$ we observe that $b_F=u_F=1$ and $H_t=\infty \times \mathbf{1}_{(\epsilon_1 \leq t)}$, with the associated copula $C_F$ being the upper Fr\'echet bound $C_F(u_1,\ldots,u_d)=\min (u_1,\ldots,u_d)$.
\end{example}

\bigskip
The following example constitutes a new parametric family of extreme-value copulas. It furthermore gives a method to approximate a distribution function with unit mean and unbounded support by one with bounded support. This can be useful for simulation purposes, see Section~\ref{subsec_finsupp}.

\bigskip
\begin{example}[Bounded support from infinite support]\label{ex_boundsupp}
Let $F$ be a distribution function with support $[0,\infty)$. Furthermore, denote $\varphi_F(\theta)\equiv \int_0^{\infty}e^{-\theta\,t}\,\mathrm{d}F(t)$ the Laplace transform of $F$. If $X$ has distribution function $F$, the random variable $X_{\theta}\equiv \{1-\exp({-\theta\,X})\}/\{1-\varphi_F(\theta)\}$ has bounded support $[0,\{1-\varphi_F(\theta)\}^{-1}]$, unit mean, and distribution function defined, for all $t \in [  0, \{ 1-\varphi_F(\theta) \} ^{-1}]$, by
\begin{gather*}
F_{\theta}(t)=F \left[  -\frac{\ln [  1-t\{1-\varphi_F(\theta)\}] }{\theta} \right].
\end{gather*}
L'Hospital's rule shows that the argument satisfies
\begin{gather*}
\lim_{\theta \searrow 0} -\frac{\ln [  1-t\{1-\varphi_F(\theta)\}] }{\theta} = t,
\end{gather*}
implying that $(F_{\theta})_{\theta>0}$ is a parametric family of distribution functions with unit mean including the original function $F$ as a marginal special case for $\theta=0$. For instance, consider a special case of Example~\ref{ex_galambos}, namely the unit exponential distribution $F(t)=1-e^{-t}$. The associated distribution function $F_{\theta}$ with bounded support is given, for all $t \in [  0, (1+\theta)/{\theta}]$, by
\begin{gather}
F_{\theta}(t) = 1- \left(  1-t\,\frac{\theta}{1+\theta} \right) ^{ {1}/{\theta}} .
\label{finsupex}
\end{gather}
It is not difficult to see that the associated L\'evy measure is
$\nu(\mathrm{d}x) = (1+\theta)\,(  1-e^{-x}) ^{\theta-1}\,e^{-x}\,\mathrm{d}x$.
\end{example}

\bigskip
\begin{example}[Exchangeable Cuadras--Aug\'e copula]
With a parameter $\theta \in [0,1]$ we consider the distribution function defined, for all $t \in [0, \infty)$, by
$F(t) = 1-\theta+\theta\times \mathbf{1}_{[1/\theta,\infty)}(t)$, 
i.e., a random variable $X$ with distribution function $F$ satisfies $\IP(X=0)=1-\theta=1-\IP(X=1/\theta)$, and in particular $\IE (X)=1$. The associated L\'evy measure is $\nu=1/\theta\times \delta_{-\ln (1-\theta)}$, and while noticing $\mathbf{1}_{[1/\theta,\infty)}(x-)=\mathbf{1}_{(1/\theta,\infty)}(x)$ we see that
\begin{gather*}
H_t = -\sum_{k \geq 1}\ln \left\{ 1-\theta+\theta\times \mathbf{1}_{( {1}/{\theta},\infty)} \left(  \frac{\epsilon_1+\cdots+\epsilon_k}{t} \right) \right\}=-\ln (1-\theta)\,\sum_{k \geq 1} \mathbf{1}_{\{\theta\,(\epsilon_1+\cdots+\epsilon_k) \leq t\}}
\end{gather*}
is a compound Poisson subordinator with intensity $1/\theta$ and constant jump sizes $-\ln (1-\theta)$. It is not difficult to see that the associated stable tail dependence function satisfies
\begin{align*}
\ell_F(t_1,\ldots,t_d)=\sum_{k=1}^{d}(1-\theta)^{d-k}\,\theta^{k-1}\,\sum_{1 \leq i_1<\cdots<i_k \leq d}\max (t_{i_1},\ldots,t_{i_k}) = \sum_{k=1}^{d}(1-\theta)^{d-k}\,t_{[k]},
\end{align*}
where $t_{[1]}\leq \cdots \leq t_{[d]}$ denotes the ordered list of the real numbers $t_1,\ldots,t_d$. The resulting one-parametric family of extreme-value copulas, defined for all $u_1, \ldots, u_d \in [0,1]$, by
\begin{gather*}
C_{\theta}(u_1,\ldots,u_d) =\prod_{k=1}^{d}u_{[k]}^{(1-\theta)^{k-1}},
\end{gather*}
corresponds to the exchangeable special case of a family first introduced in \cite{cuadras81}. Furthermore, this family falls within the larger class of L\'evy-frailty copulas studied in \cite{mai09}, which itself was shown later in \cite{mai11} to be a subfamily of Marshall--Olkin copulas, which are the survival copulas of the multivariate exponential distributions introduced in~\cite{marshall67}.
\end{example}

We end this section with a few remarks on properties of the process $(H_t)_{t \geq 0}$.

\begin{remark}[Properties of $(H_t)_{t \geq 0}$]
The stochastic process $(H_t)_{t \geq 0}$ is continuous if and only if $F$ is continuous. And this is the case if and only if the multivariate distribution of $(Y_1,\ldots,Y_d)$ is absolutely continuous for every integer $d \geq 1$. Conversely, if $F$ has a jump at $t>0$, the process $(H_t)_{t \geq 0}$ has jumps at all $t/(\epsilon_1+\cdots+\epsilon_k)$, unless it has already jumped to infinity, which can only happen if $b_F>0$.

\bigskip
If we denote by $(N_t)_{t \geq 0}$ a Poisson process with unit intensity, the considered process $(H_t)_{t \geq 0}$ may alternatively be written, for all $t \geq 0$, as
\begin{gather*}
H_t = \int_0^{\infty}-\ln  \{  F( {s}/{t}-) \}\,\mathrm{d}N_s
\end{gather*}
This shows that $(H_t)_{t \geq 0}$ falls within a family of subordinators that are strongly infinitely divisible with respect to time that is considered in Lemma~2 of \cite{bernhart15}. Generally speaking, processes of the form
\begin{gather*}
H_t = \int_0^{\infty}f( {s}/{t}) \,\mathrm{d}L_s,\quad  (L_t)_{t \geq 0} \mbox{ L\'evy subordinator}
\end{gather*}
with non-negative, left-continuous and non-increasing functions $f$ --- subject to some technical integrability conditions --- are always (right-continuous and) strongly infinitely divisible with respect to time, and hence give rise to a family of extreme-value copulas via the stochastic model (\ref{definetti_model}) according to Theorem~5.3 in \cite{maischerer13}. Whereas the article \cite{bernhart15} studies the cases $f(s)=\max (1-s,0)$ and $f(s)=\max\{-\ln (s),0\}$ but varies the L\'evy subordinator, the present article is complementary in the sense that the L\'evy subordinator is held fix (at a standard Poisson process) but the function $f(s)=-\ln \{ F(s-)\}$ is varied. 

\bigskip
As a final remark, for fixed $t \in (0, \infty)$ the random variable $H_t$ falls into a slightly more general family of stochastic representations of infinitely divisible distributions that is used as basis for random number generation in~\cite{bondesson82}.
\end{remark}

\section{Non-exchangeable extension and simulation}\label{sec_simu}

Now we assume that ${\bf F}=(F_1,\ldots,F_d)$ are possibly different distribution functions of non-negative random variables with unit mean. With a sequence of independent and identically distributed unit exponential random variables $(\epsilon_k)_{k \in \IN}$ we consider the dependent stochastic processes defined, for all $i \in \{1,\ldots,d\}$ and $t \in [0, \infty)$, by
\begin{gather*}
H^{(i)}_t\equiv -\ln \left \{\prod_{k=1}^{\infty}\,F_i \left(  \frac{\epsilon_1+\cdots+\epsilon_k}{t}- \right)  \right\},
\end{gather*}  
and define the random vector $(Y_1,\ldots,Y_d)$ by setting, for each $i \in \{ 1, \ldots, d\}$, $Y_i\equiv \inf\{t>0 : H^{(i)}_t > \xi_i\}$, where $\xi_1,\ldots,\xi_d$ are independent unit exponentially distributed random variables, independent of $(\epsilon_k)_{k \in \IN}$. The precisely same computation as in Lemma~\ref{lemma_DeFinetti} shows that, for all $t_1,\ldots,t_d \in [0, \infty)$,
\begin{gather*}
\IP(Y_1>t_1,\ldots,Y_d>t_d) = \exp\{  -\ell_{\bf F}(t_1,\ldots,t_d)\}
\end{gather*}
manifesting a non-exchangeable extension of the copula family discussed in the preceding section. In the sequel, we discuss simulation from the copula $C_{\bf F}$ associated with $\ell_{\bf F}$, which is the survival copula of the random vector $(Y_1,\ldots,Y_d)$, which has unit exponential margins.

Due to the infinite product in the definition of the processes $H^{(i)}_t$ it is not straightforward to use the stochastic model $(Y_1,\ldots,Y_d)$ in order to simulate from the extreme-value copula $C_{\bf F}$. However, if all $F_1,\ldots,F_d$ have bounded supports, i.e., $u_{F_i}<\infty$ for all $i \in \{1,\ldots,d\}$, this is possible, as shown below in Section~\ref{subsec_finsupp}. In the general case, simulation of a random vector $(U_1,\ldots,U_d)$ with distribution function $C_{\bf F}$ can be accomplished via the strategy in Algorithm~1 of \cite{dombry16}, which itself is based on an idea of Schlather~\cite{schlather02}. This algorithm is based on the random vector ${\bf Q}$ from the Pickands representation (\ref{Pickandsrepr}). More precisely, the random vector $(Z_1,\ldots,Z_d)$ with distribution function (\ref{evd}), with standard Fr\'echet margins $G_1(t)=\cdots=G_d(t)=\exp(-1/t)$ for all $t \in [0, \infty)$, has the stochastic representation
\begin{gather*}
{\bf Z}=(Z_1,\ldots,Z_d) = \left( \max_{k \geq 1}\left( \frac{Q_1^{(k)}}{\epsilon_1+\cdots+\epsilon_k}\right),\ldots,\max_{k \geq 1}\left( \frac{Q_d^{(k)}}{\epsilon_1+\cdots+\epsilon_k}\right) \right) ,
\end{gather*}
where ${\bf Q}^{(k)}$ are independent copies of ${\bf Q}$ and, independently, $(\epsilon_k)_{k \in \IN}$ is a list of independent and identically distributed exponential random variables with mean $1/d$. The simulation algorithm makes use of the fact that the components of ${\bf Q}$ are bounded from above by one, which together with the decreasingness of the sequence $\{(\epsilon_1+\cdots+\epsilon_k)^{-1}\}_{k \geq 1}$ allows to compute the involved infinite maxima as finite maxima. Concretely, we introduce the notation
\begin{align*}
{\bf Z}_n&\equiv \left( \max_{k \in\{ 1,\ldots,n\}}\left(\frac{Q_{1}^{(k)}}{\epsilon_1+\cdots+\epsilon_k}\right),\ldots,\max_{k \in\{ 1,\ldots,n\}}\left( \frac{Q_{d}^{(k)}}{\epsilon_1+\cdots+\epsilon_k}\right) \right) ,\\
m_n&\equiv \mbox{minimal component of }{\bf Z}_n
\end{align*}
for all $n \in \mathbb{N}$. Since every single component of ${\bf Q}^{(n+1)}/(\epsilon_1+\cdots+\epsilon_{n+1})$ is smaller or equal than $1/(\epsilon_1+\cdots+\epsilon_n)$,  
\begin{gather*}
{\bf Z} = {\bf Z}_{\infty} = {\bf Z}_{I},\mbox{ with } I\equiv \min\{n \in \IN : 1/(\epsilon_1+\cdots+\epsilon_n) \leq m_n\}. 
\end{gather*} 
Now $1/(\epsilon_1+\cdots+\epsilon_n)$ is almost surely decreasing to zero and $m_n$ is almost surely non-decreasing, so $I$ is almost surely finite. Consequently, in order to simulate ${\bf Z}$, it is sufficient to simulate iteratively ${\bf Z}_n$ for each successive $n=1,2,\ldots$ until the stopping criterion $n=I$ takes place, which happens almost surely in finite time. 

Apparently, when implementing this algorithm the bottleneck is the availability of a simulation algorithm for the random vector ${\bf Q}$. \cite{dombry16} demonstrate how this is possible in principle for general extreme-value copulas, and exemplarily demonstrate their general technique in case of the Gumbel and the Galambos copulas of Examples~\ref{ex_stable}--\ref{ex_galambos}. The following lemma is an application of their general idea, applied to the special case of the extreme-value copula $C_{\bf F}$.

\begin{lemma}[Pickands representation of $C_{\bf F}$] \label{lemma_pickands}
Let $F_1,\ldots,F_d$ be distribution functions of non-negative random variables with unit mean. Consider the following, mutually independent random variables:
\begin{itemize}
\item[(i)] A uniformly distributed random variable $D$ on the finite set $\{1,\ldots,d\}$.
\item[(ii)] A list of independent random variables $X_1,\ldots,X_d$ with distribution functions $F_1,\ldots,F_d$, respectively.
\item[(iii)] A list of independent random variables $M_1,\ldots,M_d$ with distribution functions $\IP(M_i \leq t) = \int_{0}^{t}x\,\mathrm{d}F_i(x)$, respectively. Note that since $F_i$ has unit mean, $x\,\mathrm{d}F_i(x)$ is a probability measure on $(0,\infty)$.
\end{itemize} 
Based on these random variables, the random vector ${\bf Q}$ associated with the extreme-value copula $C_{\bf F}$ via (\ref{relation_CF}) and (\ref{Pickandsrepr}) has the stochastic representation
\begin{gather*}
(Q_1,\ldots,Q_d) = \left(  \frac{W_1}{\sum_{i=1}^{d}W_i},\ldots,\frac{W_d}{\sum_{i=1}^{d}W_i}\right) ,
\end{gather*}
where, for each $i\in \{1,\ldots,d\}$,
\begin{gather*}
W_i = \begin{cases}
M_i & \mbox{if }i=D, \\
X_i & \mbox{if }i \neq D.
\end{cases}
\end{gather*}
\end{lemma}

\bigskip
\noindent
\textit{Proof.}
First of all, it is important to remark that the probability law $x\,\mathrm{d}F_i(x)$ does not have an atom at zero, even though $\mathrm{d}F_i(x)$ might do. This implies that the $M_i$ are strictly positive almost surely, so that the division by $W_1 + \cdots + W_d$ in the definition of ${\bf Q}$ is well-defined. Observe further that the probability distribution of the random vector ${\bf W}=(W_1,\ldots,W_d)$ is given by
\begin{gather*}
\IP({\bf W} \in \mathrm{d}{\bf x}) = \frac{1}{d}\,\sum_{j=1}^{d}x_j\,\mathrm{d}F_j(x_j)\,\prod_{i\neq j}^{d}\mathrm{d}F_i(x_i),
\end{gather*}
where ${\bf x}=(x_1,\ldots,x_d)$. Thus,
\begin{align*}
\ell_F(t_1,\ldots,t_d) & = \IE \{ \max (t_1\,X_1,\ldots,t_d\,X_d )\} = \iint_{(0,\infty)^{d}}\max (t_1\,x_1,\ldots,t_d\,x_d)\,\prod_{i=1}^{d}\mathrm{d}F_i(x_i)\\
& = \iint_{(0,\infty)^{d}}\max\left( t_1\,\frac{x_1}{\sum_{j=1}^{d}x_j},\ldots,t_d\,\frac{x_d}{\sum_{j=1}^{d}x_j} \right) \sum_{j=1}^{d}x_j\,\prod_{i=1}^{d}\mathrm{d}F_i(x_i)\\
& =d\, \iint_{(0,\infty)^{d}}\max\left( t_1\,\frac{x_1}{\sum_{j=1}^{d}x_j},\ldots,t_d\,\frac{x_d}{\sum_{j=1}^{d}x_j} \right) \frac{1}{d}\,\sum_{j=1}^{d}x_j\,\mathrm{d}F_j(x_j)\,\prod_{i\neq j}^{d}\mathrm{d}F_i(x_i)\\
& =d\, \iint_{(0,\infty)^{d}}\max\left( t_1\,\frac{x_1}{\sum_{j=1}^{d}x_j},\ldots,t_d\,\frac{x_d}{\sum_{j=1}^{d}x_j} \right) \IP({\bf W} \in \mathrm{d}{\bf x}) = d\,\IE \{ \max (t_1\,Q_1,\ldots,t_d\,Q_d ) \},
\end{align*}
completing the argument. \hfill $\Box$

\begin{figure}[h]
\centering
\includegraphics[width=0.45\linewidth]{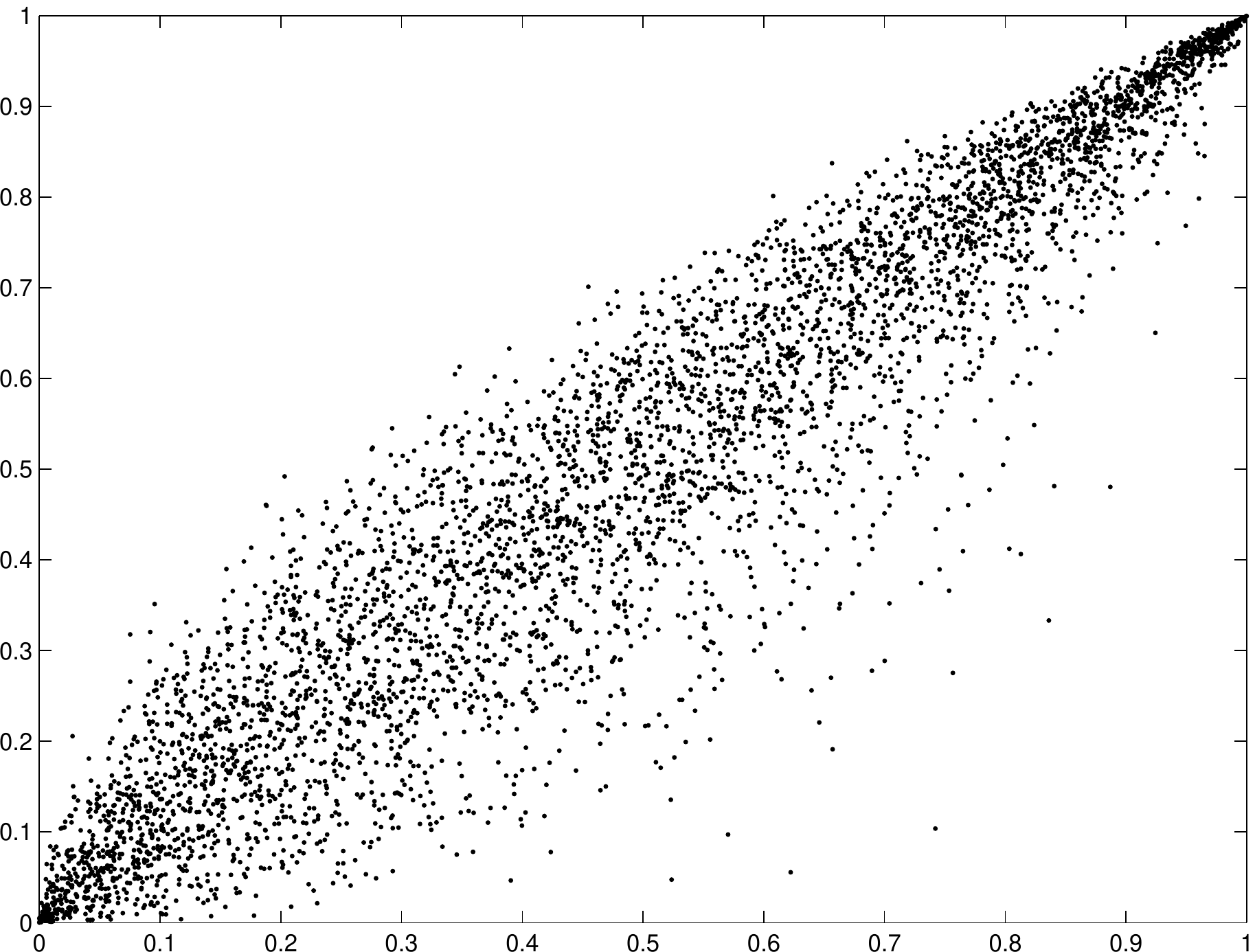}
\hfill
\includegraphics[width=0.45\linewidth]{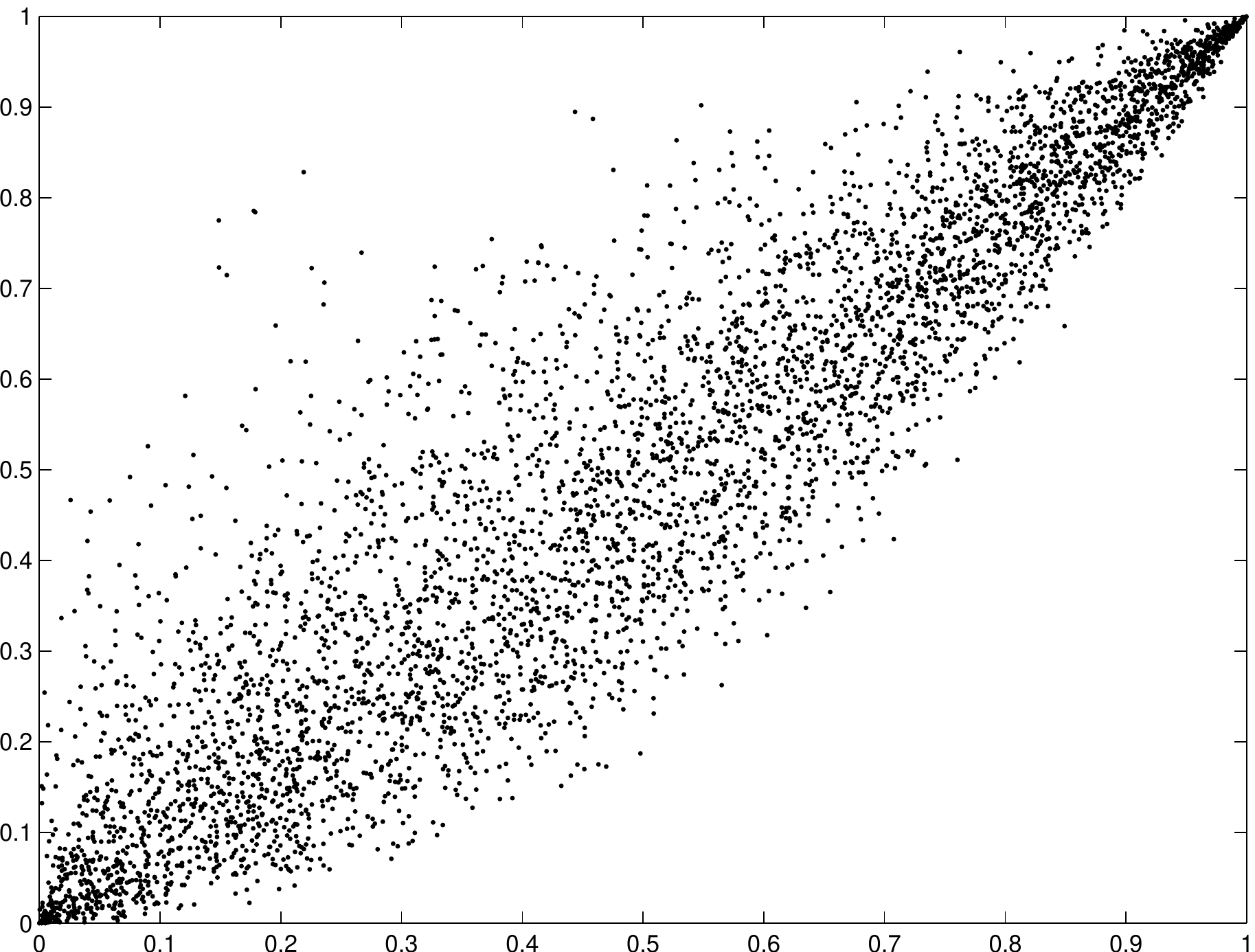}
\caption{Scatter plot of $5,000$ samples from the bivariate copula $C_{(F_1,F_2)}$. Left: $F_1$ given by (\ref{gumbeldf}) with parameter $\theta=0.1$ and $F_2$ given by (\ref{galambosdf}) with parameter $\theta=0.3$. Right: $F_1$ given by (\ref{galambosdf}) with parameter $\theta=0.1$ and $F_2$ given by (\ref{gumbeldf}) with parameter $\theta=0.3$.}
\label{fig}
\end{figure}

\bigskip
\begin{example}[When Gumbel meets Galambos in a scatter plot]
Dombry et al. \cite{dombry16} showed that if $X$ has distribution function $F$ given by (\ref{gumbeldf}) (resp.\ (\ref{galambosdf})), the random variable $M$ with probability measure $x\,\mathrm{d}F(x)$ has the stochastic representation $M \stackrel{d}{=} A^{-{\theta}}/\Gamma(1-\theta)$ (resp.\ $M \stackrel{d}{=} A^{\theta}/\Gamma(1+\theta)$), where $A$ has a $\Gamma(1-\theta,1)$-distribution (resp.\ $\Gamma(1+\theta,1)$-distribution). These stochastic representations for the random variable $M$ (together with obvious simulation algorithms via the inversion method for $X$) make Algorithm~1 in \cite{dombry16} feasible for both the Gumbel and the Galambos copula. With the help of Lemma~\ref{lemma_pickands} it is possible to simulate copulas $C_{\bf F}$ of mixed Gumbel/Galambos type.

Figure \ref{fig} visualizes scatter plots of the bivariate copula $C_{(F_1,F_2)}$ of mixed Gumbel/Galambos type. In the left plot, $F_1$ is the Fr\'echet distribution (\ref{gumbeldf}) with parameter $\theta=0.1$, and $F_2$ is the Weibull distribution (\ref{galambosdf}) with parameter $\theta=0.3$. In the right plot, we switch Gumbel and Galambos, i.e., $F_1$ is the Weibull distribution (\ref{galambosdf}) with parameter $\theta=0.1$, and $F_2$ is the Fr\'echet distribution (\ref{gumbeldf}) with parameter $\theta=0.3$.  The simulation has been accomplished by the aforementioned strategy of Algorithm~1 in \cite{dombry16} with the help of Lemma~\ref{lemma_pickands}. One observes that $C_{(F_1,F_2)}$ is not exchangeable, since the majority of points in the plot lie around a slightly skewed diagonal, in both cases skewed towards the component associated with the Weibull distribution of the Galambos case.
\end{example}

\subsection{The case of bounded supports}\label{subsec_finsupp}

The application of Algorithm~1 in \cite{dombry16} to the considered family of extreme-value copulas $C_{\bf F}$ relies on the possibility to simulate efficiently from the probability distributions $\mathrm{d}F_i(x)$ and $x\,\mathrm{d}F_i(x)$, which might not always be straightforward. Moreover, due to the great level of generality of this algorithm, which in principle is applicable to arbitrary extreme-value copulas, it is not surprising that for specific families it is possible to find alternative algorithms speeding up the simulation. In particular, the presence of a De Finetti structure is particularly well-suited for efficient simulation, especially in large dimension $d$. In general terms, this is because the latent factor, in the present case this is the sequence $(\epsilon_k)_{k \in \IN}$ in the definition of the processes $(H^{(i)}_t)_{t \geq 0}$, needs to be simulated only once. Conditioned on this simulation, $d$ independent and identically distributed random variables need to be drawn, in the present case as first passage times of the already simulated processes $(H^{(i)}_t)_{t \geq 0}$ over the trigger variates $\xi_i$. If the distribution functions $F_1,\ldots,F_d$ have bounded supports, i.e., $u_{F_i}<\infty$ for all $i \in \{1,\ldots,d\}$, the stochastic model based on the latent frailty processes $(H^{(i)}_t)_{t \geq 0}$ is shown below to be viable for this task. The bounded supports imply that the infinite products in the definition of the frailty processes become finite, so can be evaluated. If, in addition, we assume that the $F_i$ are continuous, the trigger levels $\xi_i$ are hit exactly by $(H^{(i)}_t)_{t \geq 0}$, and this hitting time $Y_i$ can be computed numerically via a bisection routine. In some special cases, this hitting time can even be computed in closed form, speeding up the simulation algorithm massively, see Example \ref{ex_simufin} below. 

As already mentioned, in this section we assume that $u_{F_i}<\infty$, $i\in \{1,\ldots,d\}$. For each $i \in \{1,\ldots,d\}$ and fixed $t \in (0, \infty)$ it follows that
\begin{gather}
H^{(i)}_t = \infty\,\times \mathbf{1}_{\{\epsilon_1 \leq b_{F_i}\,t\}}-\sum_{k=1}^{N^{(i)}_t}\ln \left\{F_i \left(  \frac{\epsilon_1+\cdots+\epsilon_k}{t}- \right) \right\},\quad N^{(i)}_t\equiv \sum_{k \geq 1}1_{\{\epsilon_1+\cdots+\epsilon_k \leq t\,u_{F_i}\}} ,
\label{finite_stochrepr}
\end{gather}
i.e., $H^{(i)}_t$ can be computed as a finite sum almost surely. According to (\ref{finite_stochrepr}), the process $(N^{(i)}_t)_{t \geq 0}$ is a Poisson process with intensity $u_{F_i}$. However, $(H^{(i)}_t)_{t \geq 0}$ is obviously not a compound Poisson process, since the jump sizes depend on both time $t$ and the inter-arrival times of $N^{(i)}_t$. Instead, it resembles a shot-noise process. 

If the $F_i$ are all continuous, exact simulation of $(Y_1,\ldots,Y_d)$ is possible according to Algorithm~\ref{algo_finsupp} below, which is briefly explained. Introducing the notation $ H^{(i,0)}_t \equiv \infty\times \mathbf{1}_{(\epsilon_1<b_{F_i}\,t)}$, the stochastic representation (\ref{finite_stochrepr}) shows that, for all $k \in \mathbb{N}$,
\begin{gather*}
H^{(i)}_t = H^{(i,0)}_t+H^{(i,1)}_t+\cdots+H^{(i,N^{(i)}_t)}_t,\quad H^{(i,k)}_t = -\ln \left\{F_i \left(  \frac{\epsilon_1+\cdots+\epsilon_k}{t}- \right) \right\},
\end{gather*}
where the stochastic process $(H^{(i,k)}_t)_{t \geq 0}$ is identically zero on $[ 0,(\epsilon_1+\cdots+\epsilon_k)/u_{F_i} ) $. Consequently, for any integer $N$ the computation of a path of $(H^{(i)}_t)_{t \geq 0}$ until $t=(\epsilon_1+\cdots+\epsilon_N)/u_{F_i}$ requires only a sum with $N-1$ summands, and at the final time $t=(\epsilon_1+\cdots+\epsilon_N)/u_{F_i}$, the process $(H^{(i)}_t)$ takes the value
\begin{gather*}
x^{(i)}_N\equiv H^{(i)}_{\frac{\epsilon_1+\cdots+\epsilon_N}{u_{F_i}}} =\infty\times \mathbf{1}_{\{{\epsilon_1}/{(\epsilon_1+\cdots+\epsilon_N)}<{b_{F_i}}/{u_{F_i}}\}} -\sum_{k=1}^{N-1}\ln \left\{F_i \left(  \frac{\epsilon_1+\cdots+\epsilon_k}{\epsilon_1+\cdots+\epsilon_N}\,u_{F_i}- \right) \right\}.
\end{gather*}
For each $i \in \{1,\ldots,d\}$, denoting the minimal index $N$ such that $x^{(i)}_N>\xi_i$ by $I(i)$, the first-passage time $Y_i$ lies between $(\epsilon_1+\cdots+\epsilon_{I(i)})/u_{F_i}$ and $(\epsilon_1+\cdots+\epsilon_{I(i)+1})/u_{F_i}$, and on that interval the process $(H^{(i)}_t)$ takes the form
\begin{align*}
H^{(i)}_t = H^{(i,1)}_t+\cdots+H^{(i,I(i))}_t=\infty\times \mathbf{1}_{\{\epsilon_1<b_{F_i}\,t\}}-\sum_{k=1}^{I(i)}\ln \left\{F_i \left(  \frac{\epsilon_1+\cdots+\epsilon_k}{t}- \right) \right\}.
\end{align*}

The resulting simulation algorithm is given in pseudo code as follows. 

\begin{algorithm}[Exact simulation of $C_{\bf F}$ in case of bounded supports and continuous $F_i$] \label{algo_finsupp}
We assume that $F_1,\ldots,F_d$ have bounded supports $u_{F_1},\ldots,u_{F_d}<\infty$ and are continuous. For fixed integer $d \in \mathbb{N}$, we simulate $(U_1,\ldots,U_d)$ with distribution function $C_{\bf F}$.
\begin{itemize}
\item[(1.1)] Draw $\xi_1,\ldots,\xi_d$ independent and identically distributed with unit exponential law.
\item[(1.2)] Initialize $N\equiv 1$, draw a unit exponentially distributed random variable $\epsilon_N$, initialize a list object $S\equiv [\epsilon_N]$.
\item[(1.3)] Initialize $I\equiv (0,\ldots,0)$ and $x\equiv (0,\ldots,0)$, two $d$-dimensional random vectors with all entries zero.
\item[(2)] While $x(i)<\xi_i$ for at least one $i \in \{1,\ldots,d\}$, perform the following steps:
\begin{itemize}
\item[(2.1)] Set $N\equiv N+1$, draw a unit exponentially distributed random variable $\epsilon_{N}$, and set $S\equiv [S,\,S(N-1)+\epsilon_{N}]$.
\item[(2.2)] For $i\in \{1,\ldots,d\}$, compute $x(i)\equiv \infty\times \mathbf{1}_{ \{ {S(1)}/{S(N)}< {b_{F_i}}/{u_{F_i}} \}}-\sum_{k=1}^{N-1}\ln \Big[ F_i\{  \frac{S(i)}{S(N)}\,u_{F_i}\} \Big]$.
\item[(2.3)] For $i\in \{1,\ldots,d\}$, set $I(i)\equiv N-1$, if $x(i)>\xi_i$ and $I(i)=0$.
\end{itemize}
\item[(3)] Return $(U_1,\ldots,U_d)=(\exp(-Y_1),\ldots,\exp(-Y_d))$, where $Y_i$ is the unique root of the function
\begin{gather*}
f_i(t)\equiv \infty \times \mathbf{1}_{\{\epsilon_1<b_{F_i}\,t\}}-\sum_{k=1}^{I(i)}\ln  \{ F_i(  {S(k)}/{t}) \} - \xi_i,
\end{gather*}
in the interval $( {S\{I(i)\}}/{u_{F_i}}, {S\{I(i)+1\}}/{u_{F_i}}] $, for $i\in \{1,\ldots,d\}$. This root search may be accomplished numerically via a bisection routine, or in special cases even in closed form, see Example~\ref{ex_simufin} below.
\end{itemize}
\end{algorithm}

\medskip
\begin{example}[Comparison of simulation algorithms in case of bounded support and continuous $F$]\label{ex_simufin}
Consider the case $F_1=\cdots=F_d=F$ for the distribution function $F(t)=t/2$ with $u_F=2$ (i.e., $\theta=1$ in the parametric family (\ref{finsupex})). Due to the very simple structure of $F$ in that particular case, the root search in Step (3) of Algorithm~\ref{algo_finsupp} can be solved in closed form, yielding, for each $i \in \{ 1, \ldots, d\}$,
\begin{gather*}
Y_i = \frac{1}{2}\,\exp \left[  \frac{1}{I(i)}\,\left[  \xi_i+\ln \left\{  \prod_{k=1}^{I(i)}(\epsilon_1+\cdots+\epsilon_k) \right\} \right] \right] ,
\end{gather*}
where $I(i)$ is the minimal natural number $N$ for which 
\begin{gather*}
x_N=-\sum_{k=1}^{N-1}\ln \left(  \frac{\epsilon_1+\cdots+\epsilon_k}{\epsilon_1+\cdots+\epsilon_N} \right) >\xi_i,
\end{gather*}
for all $i \in \{1,\ldots,d\}$. The resulting simulation algorithm is much faster than Algorithm~1 in \cite{dombry16} for this particular family, especially in large dimensions, because it makes use of the (dimension-free) De Finetti structure. Table~\ref{tab} shows the CPU time required for the generation of $5,000$ samples from the $d$-dimensional copula $C_{F}$ in \texttt{MATLAB} on a standard PC. For the implementation of Algorithm~1 in \cite{dombry16}, the required simulation algorithms for the random variables $X_i$ and $M_i$ according to Lemma~\ref{lemma_pickands} have been implemented via the inversion method, i.e., 
$$
X_i \stackrel{d}{=}2\,U, \quad M_i \stackrel{d}{=}2\,\sqrt{U}
$$ 
for $U$ uniform on $[0,1]$. It is observed that the simulation strategy based on the De Finetti structure is hardly affected when the dimension $d$ is increased, which is not the case for the general algorithm based on the Pickands measure. 
\end{example}

\begin{table}[!htbp]
\begin{center}
\begin{tabular}{lcccccc}
Dimension $d$ & $2$ & $5$ & $10$ & $25$ & $50$ & $100$\\
\hline
Simulation based on De Finetti& $0.56$ & $0.72$ & $0.89$ & $1.08$ & $1.25$ & $1.50$\\
Simulation based on Pickands measure & $0.51$ & $1.59$ & $3.81$ & $12.90$ & $30.92$ & $79.98$\\
\hline
\end{tabular}
\end{center}
\caption{CPU time in seconds, required for the generation of $5,000$ samples from the $d$-dimensional copula $C_F$ with $F(t)=t/2$ for all $t \in [0,2]$, implemented in \texttt{MATLAB} on a standard PC. The first row corresponds to Algorithm \ref{algo_finsupp}, the second row to \cite[Algorithm 1]{dombry16} with the help of Lemma~\ref{lemma_pickands}.}
\label{tab}
\end{table}%

\section{Conclusion} \label{sec_conc}
The family of extreme-value copulas whose associated stable tail dependence function is the expected scaled maximum of independent, non-negative random variables with distribution functions $F_1,\ldots,F_d$ has been investigated. In the exchangeable case $F_1=\cdots=F_d$ a stochastic representation in which the components are conditionally independent and identically distributed in the sense of De Finetti's Theorem has been derived and explored. Furthermore, it has been demonstrated how the De Finetti structure can be used for efficient simulation. Especially in large dimensions the method has been demonstrated to outperform a more general simulation scheme of \cite{dombry16}, which has been recalled and applied to the current setting.

\section*{Acknowledgments}
Helpful comments on an earlier version of the paper by two anonymous referees are gratefully acknowledged.

\end{document}